\documentclass[publicdomain]{eptcs}
\providecommand{\event}{CompMod 2009} 
\providecommand{\volume}{??}
\providecommand{\anno}{20??}
\providecommand{\firstpage}{1}
\providecommand{\eid}{??}

\usepackage{breakurl}        
\usepackage{graphicx}
\usepackage{amsfonts}
\usepackage{amssymb}
\usepackage{amsmath}

\newcommand{\XX}{\mathbf{X}}
\newcommand{\xx}{\mathbf{x}}
\newcommand{\nn}{\mathbf{\nu}}

\title{On the Interpretation of Delays in Delay Stochastic Simulation
of Biological Systems}

\author{Roberto~Barbuti \and Giulio~Caravagna \and Andrea Maggiolo-Schettini \and  Paolo~Milazzo
\institute{Dipartimento di Informatica, Universit\`a di Pisa\\
Largo Pontecorvo 3, 56127 Pisa, Italy.\\}
\email{\{barbuti,caravagn,maggiolo,milazzo\}@di.unipi.it}
}

\def\titlerunning{On the Interpretation of Delays in Delay Stochastic Simulation
of Biological Systems}
\def\authorrunning{R. Barbuti \emph{et al.}}
\begin{document}
\maketitle

\begin{abstract}
Delays in biological systems may be used to model events for which the
underlying dynamics cannot be precisely observed. Mathematical modeling of
biological systems with delays is usually based on Delay Differential Equations
(DDEs), a kind of differential equations in which the derivative of the unknown
function at a certain time is given in terms of the values of the function at
previous times. In the literature, delay stochastic simulation algorithms  have
been proposed. These algorithms follow a ``delay as duration'' approach, namely
they are based on an interpretation of a delay as the elapsing time between the
start and the termination of a chemical reaction. This interpretation is not
suitable for some classes of biological systems in which species involved in a
delayed interaction can be involved at the same time in other interactions. We
show on a DDE model of tumor growth that the delay as duration approach for
stochastic simulation is not precise, and we propose a simulation algorithm
based on a ``purely delayed'' interpretation of delays which provides better
results on the considered model.
\end{abstract}

\section{Introduction}

Biological systems can often be modeled at different abstraction levels. A
simple event in a model that describes the system at a certain level of detail
may correspond to a rather complex network of events in a lower level
description. The choice of the abstraction level of a model usually depends on
the knowledge of the system and on the efficiency of the analysis tools to be
applied to the model.

Delays may appear in models of biological systems at any abstraction level, and are
associated with events whose underlying dynamics either cannot be precisely
observed or is too complex to be handled efficiently by analysis tools.
Roughly, a delay may represent the time necessary for the underlying network of
events to produce some result observable in the higher level model.

Mathematical modelling of biological systems with delays is mainly based on
delay differential equations (DDEs), a kind of differential equations in which
the derivative of the unknown function at a certain time is given in terms of
the values of the function at previous times. In particular, this framework is
very general and allows both simple (constant) and complex (variable
or distributed) forms of delays to be modeled.

As examples of DDE models of biological systems we
mention \cite{BERETTA,ZHANGA,MARTIN,VR03,HIV}. In~\cite{BERETTA,ZHANGA} an epidemiological model
is defined that computes the theoretical number of people
infected with a contagious illness in a closed population over time; in the
model a delay is used to model the length of the infectious period.
In~\cite{MARTIN} a simple predator-prey model with harvesting and time delays is
presented; in the model a constant delay is used based on the assumption that
the change rate of predators depends on the number of prey and predators at some
previous time. Finally, models of tumor growth~\cite{VR03} and of HIV cellular
infection~\cite{HIV} have been presented and analyzed by using DDEs.

Models based on DDEs, as their simplest versions based on ordinary differential
equations (ODEs), may be studied either analytically (by finding the solution
of the equations, equilibria and bifurcation points) or via approximated numerical solutions.
However, for complex real models analytical solutions are often difficult or
impossible to be computed, whereas their approximated numerical solution is more feasible.

Models based on differential equations, although very useful when dealing with
biological systems involving a huge number of components, are not suitable to
model systems in which the quantity of some species is small. This is caused
by the fact that differential equations represent discrete quantities with
continuous variables, and when quantities are close to zero this becomes a too
imprecise approximation. In these cases a more precise description of systems
behaviour can be obtained with stochastic models, where quantities are discrete
and stochastic occurrence of events is taken into account.

The most common analysis technique for stochastic models is stochastic
simulation that, in the case of models of biological systems without delays,
often exploits Gillespie's Stochastic Simulation Algorithm (SSA) of chemical
reactions \cite{G77}, or one of its approximated variants \cite{G01,CGP05}. In
recent years, the interest for stochastic delayed processes increased~\cite{SW08}.
In~\cite{BARRIO} a Delay Stochastic Simulation Algorithm (DSSA)
has been proposed,  this algorithm gives an interpretation as durations to delays.
The delay associated
with a chemical reaction whose reactants are consumed (i.e. are not also
products) is interpreted as the duration of the reaction itself. Such an
interpretation implies that the products of a chemical reaction with a delay are
added to the state of the simulation not at the same time of reactants
removal, but after a quantity of time corresponding to the delay. Hence,
reactants cannot be involved in other reactions during the time modeled by the
delay.

We argue that the interpretation of delay as duration is not always suitable
for biological systems. We propose a simple variant of
the DSSA in which reactants removal and products insertion are performed
together after the delay. This corresponds to a different interpretation of
delays, that is the delay is seen as the time needed for preparing an event
which happens at the end of the delay. An example of a biological behavior which can be
suitably modelled by this interpretation is mitosis. Cell mitosis is characterized
by a pre--mitotic phase and by a mitotic phase (cell division). The pre--mitotic phase
prepares the division of the cell, when a cell undergoes the mitotic process, the
pre--mitotic phase can be seen as a delay before the real cell division. During the pre--mitotic
phase the cell can continue to interact with the environment, for example it can die.
The DSSA in~\cite{BARRIO} cannot model this interactions because the reactants (in this case
the cell itself) are removed at the beginning of reaction and the products are added at
its end (that is after the delay).

In this paper we start by recalling the definition of DDEs and a DDE model of
tumor growth \cite{VR03}. Then, we give a stochastic model of the considered tumor
growth example and simulate it by using the DSSA introduced in~\cite{BARRIO} and
based on an interpretation of delays as durations. Finally, we propose a new
interpretation of delays and, consequently, a new variant of the DSSA that we apply to the
considered tumor growth example. At the end of the paper we discuss further improvements of our approach and we draw some conclusions.

\section{Delay Differential Equations (DDEs)}

The mathematical modeling of biological systems is often based on Ordinary
Differential Equations (ODEs) describing the dynamics of the considered systems
in terms of variation of the quantities of the involved species over time.

Whenever phenomena presenting a delayed effect are described by differential equations, we move from
ODEs to \textit{Delay Differential Equations} (DDEs). In DDEs the derivatives at current time depend on some past states of the system. The general form of a DDE for $X(t)\in \mathbb{R}^n$ is
\[
    \frac{dX}{dt}=f_x(t,X(t),X_t),
\]
where $X_t=\{X(t'):t'\leq t\}$ represents the trajectory of the solution in the
past.

The simplest form of DDE considers \emph{constant delays}, namely consists of
equations of the form
\[
\frac{dX}{dt}=f_x(t,X(t),X(t-\sigma_1),\ldots,X(t-\sigma_n))
\]
with $ \sigma_1>\ldots>\sigma_n\geq 0$ and $\sigma_i \in \mathbb{R}$. This form
of DDE allows models to describe events having a fixed duration. They have been
used to describe biological systems in which events have a non-negligible
duration \cite{BERETTA,ZHANGA,MARTIN} or in which a sequence of
simple events is abstracted as a single complex event associated with a duration
\cite{VR03,HIV}.

In what follows we recall an example of DDE model of a biological system that
we shall use to compare delay stochastic simulation approaches.

\subsection{A DDE model of tumor growth}\label{sect:dde-model}

Villasana and Radunskaya proposed in \cite{VR03} a DDE model of tumor growth
that includes the immune system response and a phase-specific drug able to
alter the natural course of action of the cell cycle of the tumor cells.

The cell cycle is a series of sequential events leading to cell replication via cell division.
It consists of four phases: G$_1$, S, G$_2$ and M. The first three phases (G$_1$, S, G$_2$)
are called interphase. In these phases, the main event which happens is the replication of DNA.
In the last phase (M), called mitosis, the cell segregates the duplicated sets of chromosomes between daughter
cells and then divides. The duration of the cell cycle  depends on the type of cell (e.g
a human normal cell takes approximately 24 hours to perform a cycle).

The model in \cite{VR03} considers three populations of cells: the immune system, the population of
tumor cells during cell cycle interphase, and the population of tumor cells
during mitosis. A delay is used to model the duration of the interphase, hence
the model includes a delayed event that is the passage of a tumor cell from the
population of those in the interphase to the population of those in the mitotic
phase.
In the model the effect of a phase-specific drug, able to arrest tumor cells
during the mitosis, is studied. Such a drug has a negative influence also on the
survival of cells of the immune system.

In this paper we study a simplified version of the model (presented in subsection 4.1.2 of \cite{VR03}), where the effects of the immune response and  of the drug are not taken into account.
The simplified model, which considers only tumor cells (both in pre-mitotic and mitotic phases), consists of the following DDEs:
\begin{align*}
  \frac{dT_I}{dt} & = 2a_4T_M - d_2 T_I - a_1 T_I(t-\sigma)&& T_I(t) = \phi_0(t) \mbox{ for } t \in [-\sigma, 0]\\
  \frac{dT_M}{dt} & = a_1T_I(t-\sigma) - d_3 T_M - a_4 T_M&& T_M(t) = \phi_1(t) \mbox{ for } t \in [-\sigma, 0]
\end{align*}
Function $T_I(t)$ denotes the population of tumor cells during interphase at
time t, and function $T_M(t)$ denotes the tumor population during mitosis at time $t$.
The terms $d_2T_I$ and $d_3 T_M$ represent  cell deaths, or apoptoses. The
constants  $a_1$ and $a_4$ represent the phase change rates from interphase to
mitosis ($a_1$) and back ($a_4$). In the following we shall denote with
$d$ the rate at which mitotic cells disappear, namely $d =  d_3 + a_4 $.

We assume that cells reside in the interphase at least $\sigma$
units of time; then the number of cells that enter mitosis at time $t$ depends
on the number of cells that entered the interphase $\sigma$ units of time before.
This is modeled by the terms $T_I (t - \sigma)$ in the DDEs. Note that each cell
leaving the mitotic phase produces two new cells in the $T_I$ population (term
$2a_4T_M$).
In the model the growth of
the tumor cell population is obtained only through mitosis, and is given by the constants
$a_1$, $a_4$, and $\sigma$ which regulate the pace of cell division.
The delay $\sigma$ requires the values of $T_I$ and $T_M$ to be given also in the
interval $[-\sigma, 0]$: such values are assumed to be constant in the considered
interval, and hence equal to the values of $T_I$ and $T_M$ at time $0$.

\begin{figure}[t]
\begin{center}
 \includegraphics[width=10cm]{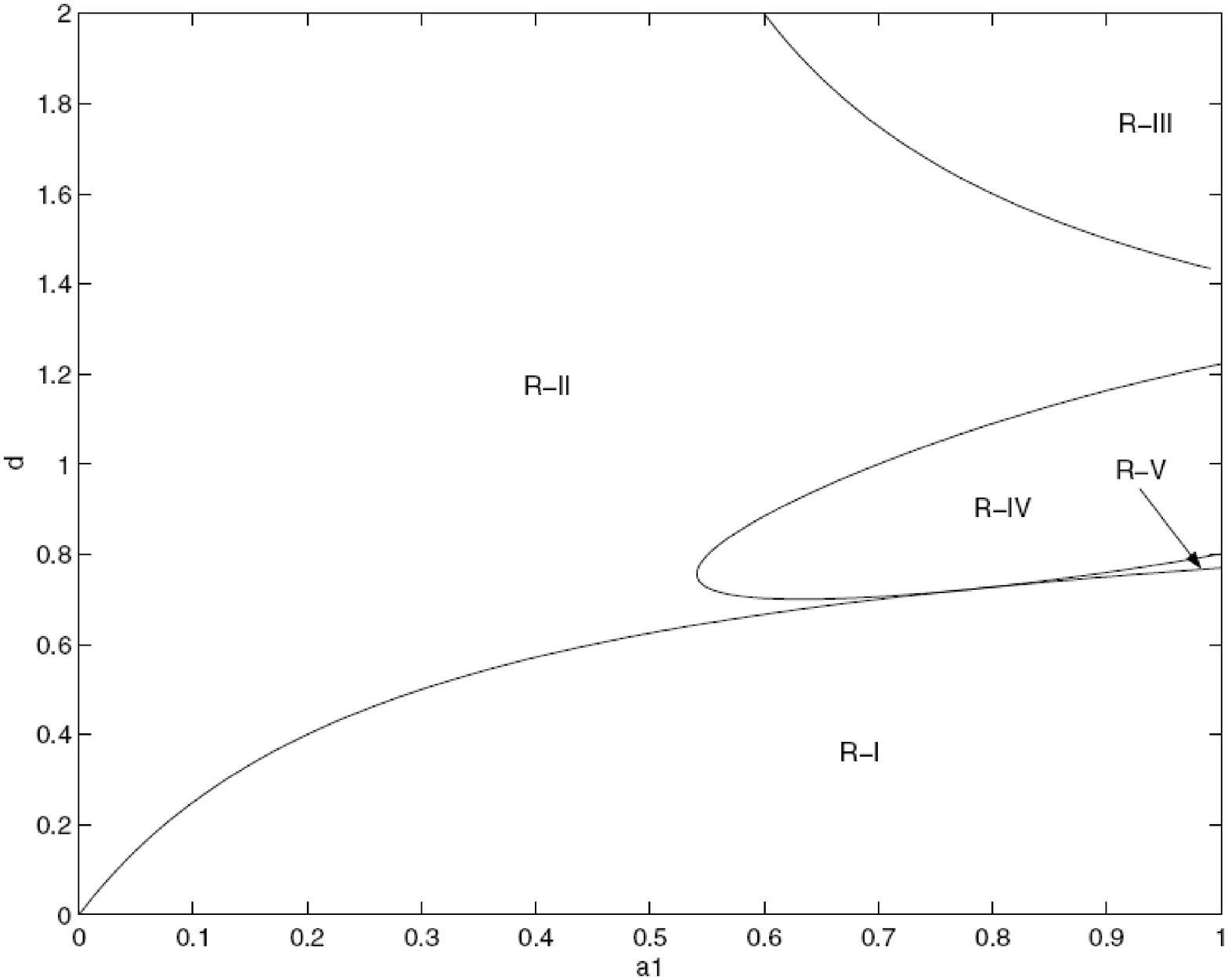}
 \caption{The regions which describe the different behaviours of the DDE model
by varying parameters $a_1$ and $d$ (picture taken from \cite{VR03}).}\label{fig:regions}
\end{center}
\end{figure}

The analytic study of the DDEs constituting the model gives $(0,0)$ as unique
equilibrium. In Figure \ref{fig:regions} (taken from \cite{VR03}) some results
are shown of the study of the model by varying $a_1, d$ and $\sigma$ and by
setting the parameters $a_4$ and $d_2$ to $0.5$ and $0.3$, respectively. Figure
\ref{fig:regions} shows five regions.

When $\sigma = 0$, the region in which the tumor grows is R-I, while in the other
regions the tumor decays.

\begin{figure} [p]
\begin{center}
 \includegraphics[width=11cm]{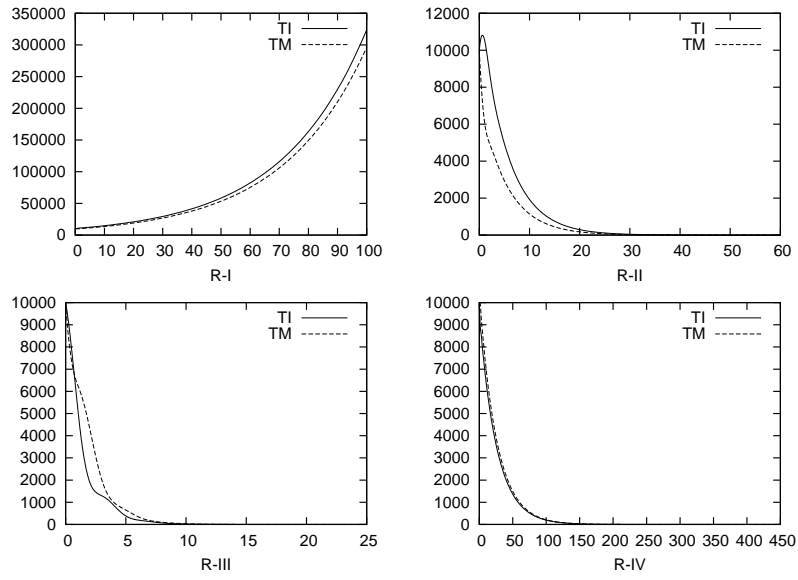}
 \caption{Results of the numerical solution of the DDE model with $\sigma=1$ for the regions described in Figure~\ref{fig:regions}. On the x-axis time is given in {\em days} and on the y-axis is given the {\em number of cells}. }\label{fig:DDE-1}
\end{center}
\end{figure}

\begin{figure}[p]
\begin{center}
 \includegraphics[width=11cm]{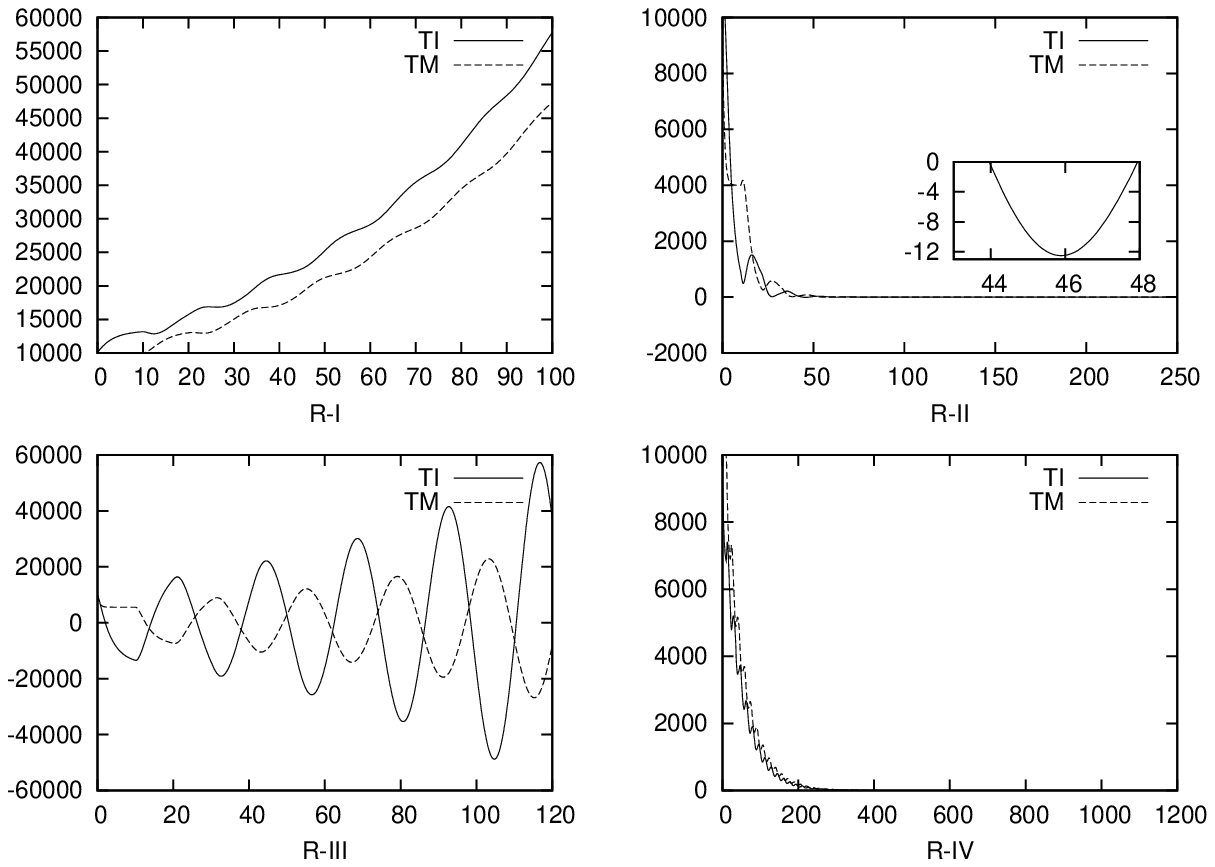}
 \caption{Results of the approximated numerical simulation of the DDE model with $\sigma=10$ for the regions described in
 Figure~\ref{fig:regions}. On the x-axis time is given in {\em days} and on the y-axis is given the {\em number of cells}.}\label{fig:DDE-10}
\end{center}
\end{figure}

When the delay is present ($\sigma > 0$), the growth region is essentially
unaltered, but the decay is split in regions in which the tumor has  different
behaviours: in regions R-II $\cup$ R-IV the tumor still decays, but in regions
R-III $\cup$ R-V, when the value of $\sigma$ is sufficiently large, the equilibrium
becomes unstable. This is shown in Figures \ref{fig:DDE-1} and
\ref{fig:DDE-10}.

Figure  \ref{fig:DDE-1} describes the behaviour of the model, obtained by numerical solutions, inside the regions R-I, R-II, R-III, and R-IV, when $\sigma = 1$. Actually, we considered the point $(0.6,0.6)$ in R-I,
the point $(0.4,1.0)$ in R-II, the point $(1.0,1.8)$ in R-III, the point $(0.8,0.8)$ in R-IV and
an initial state consisting in $10^5$ tumor cells in the interphase and $10^5$ tumor cells in mitosis.
We shall use always this parameters in the rest of the paper. In the figure, we can observe that, while the tumor grows in region R-I, it decays in all the others.

Figure  \ref{fig:DDE-10} describes the behaviour of the model when $\sigma = 10$.
In regions R-I and R-IV the tumor has the same behaviour as before. In region
R-II it decays after some oscillations, while in region R-III it expresses an
instability around the equilibrium. However, remark that values of $T_M$ and $T_I$
under $0$ are not realistic, and, as we will see in the following, they
cannot be obtained by stochastic simulations.

\section{Delay Stochastic Simulation}

In this section we present algorithms for the stochastic simulation of biological systems with delays. Firstly, we introduce a well--known formulation of one of these algorithms and we analyze the results of the simulations of the stochastic model equivalent to the one presented in the previous section. Secondly, we propose a variant of this algorithm and we compare the results of the simulations done by using this algorithm with those of the simulation done by using the original one.

All the simulations and the algorithms that we are going to present in this section have been implemented in
the software tool \texttt{Delay Sim}. This tool, avalaible at \texttt{http://www.di.unipi.it/msvbio}, has been written in Java.

\subsection{The Delay as Duration Approach (DDA)}

In~\cite{BARRIO} Barrio \textit{et al.} introduced a \textit{Delay Stochastic Simulation Algorithm} (DSSA)
by adding delays to Gillespie's Stochastic Simulation Algorithm (SSA)~\cite{G77}. The algorithm has been used to explain more carefully than with DDE models the observed sustained oscillations in the expression levels of some proteins.

In order to recall the definition of the algorithm in~\cite{BARRIO} we consider a well--stirred system of \textit{molecules} of $N$ chemical \textit{species} $\{S_1, \ldots,S_N\}$
interacting through $M$ chemical \textit{reaction channels} $\mathcal{R}={R_1, \ldots, R_M}$. We assume the volume and the temperature of the system to be constant. We denote the number of molecules of species $S_i$ in the system at time $t$ with $X_i(t)$, and we want to study the evolution of the \textit{state vector} $\XX(t)=(X_1(t), \ldots, X_N(t))$, assuming that the system was initially in some state $\XX(t_0)=\mathbf{x}_0$.

A reaction channel $R_j$ is characterized mathematically by three quantities. The first is its \emph{state--change vector} $\nn_j=(\nu_{1j},\ldots,\nu_{Nj})$, where $\nu_{ij}$ is defined to be the change in the $S_i$ molecular population caused by one $R_j$ reaction; let us denote each state--change vector $\nu_j$ as a the composition of the state--change vector for reactants, $\nu_j^r$, and the state--change vector for products, $\nu_j^p$, noting that $\nu_j = \nu_j^r + \nu_j^p$. For instance, given two species $A$ and $B$, a reaction of the form $A\xrightarrow{} B$ is described by the vector of reactants $(-1,0)$, by the vector of products $(0,1)$ and by the state--change vector $(-1,1)$; differently, a reaction of the form $A\xrightarrow{} A+B$ is described by the vector of reactants $(-1,0)$, by the vector of products $(1,1)$, and by the state--change vector $(0,1)$.

The second characterizing quantity for a reaction channel $R_j$ is its \emph{propensity function} $a_j(\xx)$; this is defined, accordingly to~\cite{G77}, so that, given $\XX(t)=\xx$, $a_j(\xx)dt$ is
the probability of reaction $R_j$ to occur in state $\xx$ in the time interval $[t,t+dt]$. As stated in~\cite{G77}, the probabilistic definition of the propensity function finds its justification in physical theory.

The other characterizing quantity is a constant delay defined by a real number $\sigma \geq 0$. Following Barrio \textit{et al.}, we classify reactions with delays into two categories: non-consuming reactions, where the reactants are also products (e.g. $A\xrightarrow{} A+B$), and consuming reactions, where some of the reactants are consumed (e.g. $A\xrightarrow{} B$). Throughout the paper, we denote the set of non-consuming reactions with delay by $\mathcal{R}_{nc}$, the set of consuming reactions with delay by  $\mathcal{R}_{c}$, and the reactions without delays by $\mathcal{R}_{nd}$; notice that $\mathcal{R}=\mathcal{R}_{nc} \cup \mathcal{R}_{c} \cup \mathcal{R}_{nd}$.

\begin{figure}[t]
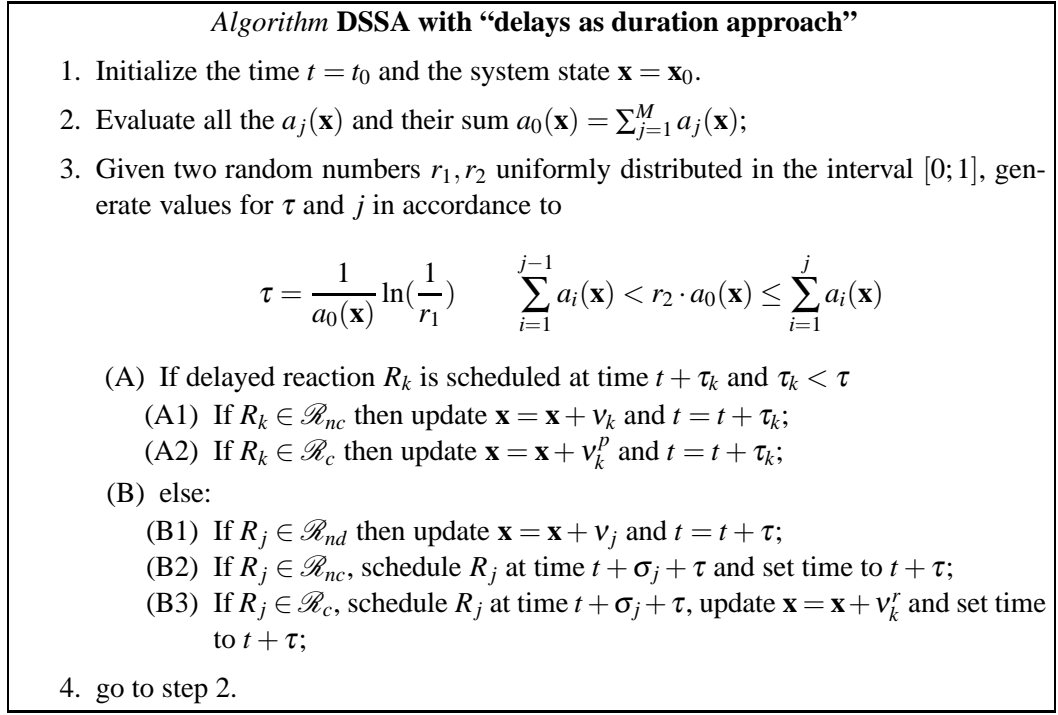

\framebox{
\begin{minipage}{0.85\textwidth}
\begin{center}
\noindent \textit{Algorithm} \textbf{DSSA with ``delays as duration approach''}
\begin{enumerate}
\item Initialize the time $t = t_0$ and the system state $\xx=\xx_0$.

\item Evaluate all the $a_j(\xx)$ and their sum $a_0(\xx)=\sum_{j=1}^Ma_j(\xx)$;

\item Given two random numbers $r_1,r_2$ uniformly distributed in the interval $[0;1]$, generate values for $\tau$ and $j$ in accordance to
{\[
 \tau = \dfrac{1}{a_0(\xx)} \ln(\dfrac{1}{r_1}) \qquad \sum_{i=1}^{j-1} a_i(\xx) < r_2 \cdot a_0(\xx) \leq \sum_{i=1}^j  a_i(\xx)
\]}
\begin{enumerate}
 \item[(A)] {If delayed reaction $R_k$ is scheduled at time $t+\tau_k$ and $\tau_k < \tau$}
 \begin{enumerate}
 \item[(A1)] {If $R_k\in \mathcal{R}_{nc}$  then update $\xx = \xx + \nu_k$ and $t = t + \tau_k$;}
 \item[(A2)] {If $R_k\in \mathcal{R}_{c}$  then update $\xx = \xx + \nu_k^p$ and $t = t + \tau_k$;}
\end{enumerate}

 \item[(B)] {else:}
 \begin{enumerate}
 \item[(B1)] {If $R_j \in \mathcal{R}_{nd}$ then update $\xx = \xx + \nu_j$ and $t = t + \tau$;}
 \item[(B2)]  {If $R_j \in \mathcal{R}_{nc}$, schedule $R_j$ at time $t+\sigma_j+\tau$ and set time to $t+\tau$};
 \item[(B3)]  {If $R_j \in \mathcal{R}_{c}$, schedule $R_j$ at time $t+\sigma_j+\tau$, update $\xx = \xx + \nu_k^r$ and set time to $t+\tau$};
\end{enumerate}
\end{enumerate}
\item go to step $2$.
\end{enumerate}
\end{center}
\end{minipage}}
\caption{The DSSA with ``delays as duration approach'' proposed in~\cite{BARRIO}.}
\label{fig:DSSA-DDA}
\end{figure}

By adding delays to the SSA, Barrio \textit{et al.} provide a method to model the firing of a reaction with delay based on the previously given classification. Formally, given a system in state $\XX(t)=\xx$, let us denote with $\tau$ the stochastic time quantity computed as in the SSA representing the putative time for next reaction to fire. Let us assume to choose to fire a non-consuming reaction with delay (a reaction from set $\mathcal{R}_{nc}$); then the reaction is scheduled at time $t+\sigma+\tau$ where $\sigma$ is the delay of the reaction. Furthermore, the clock is increased to the value $t+\tau$ and the state does not change.
On the contrary, if a consuming reaction with delay (a reaction from set $\mathcal{R}_{c}$) is chosen to fire, then its reactants are immediately removed from the state $\xx$, the insertion of the products is scheduled at time $t+\sigma+\tau$, and, finally, the clock is increased to the value $t+\tau$. Reactions from set $\mathcal{R}_{nd}$ (non--delayed reactions) are dealt with exactly as in the SSA. The DSSA by Barrio \textit{et al.} is given in Figure~\ref{fig:DSSA-DDA}.

We discuss now on the scheduling of the reactions with delay. When a non-consuming reaction is chosen, the algorithm does not change state, but simply schedules the firing of the reaction at time $t+\sigma_j+\tau$ (step $(B2)$). The reaction will complete its firing (reactants and products will be removed and inserted, respectively) when performing steps $(A)$ and $(A1)$.

Differently, as regards consuming reactions, the removal of the reactants is done at time instant $t$  (step $(B3)$) preceding the time instant of insertion of the products (steps $(A)$ and $(A2)$), namely the time at which the insertion is scheduled, $t+\sigma_j+\tau$. Notice that the removed reactants cannot have other interactions during the time interval $[t,t+\sigma_j+\tau)$.

As the reactants cannot have other interactions in the time quantity passing between the removal of the reactants and the insertion of the products, then this quantity can be seen as a duration needed for the reactants to exclusively complete the reaction. Since the approach of Barrio {\em at al.} gives this interpretation of delays we shall call it ``delays as duration approach''(DDA).

As regards the handling of the scheduled events (step $(A)$ of the algorithm), if in the time interval $[t;t+\tau)$ there are scheduled reactions, then $\tau$ is rejected and the scheduled reaction is handled. Since generating random numbers is a costly operation, other authors defined variants of the DSSA that avoid rejecting $\tau$ in the
handling of scheduled reactions~\cite{CAI,A08}. However, the interpretation of the delays used to define these variants is the same as that of Barrio {\em et al.}.

This interpretation of delays may not be precise for all biological systems. In particular, it may be not precise if in the biological system the reactants can have other interactions during the time window modeled by the delay. The tumor growth system we have recalled in Section~\ref{sect:dde-model} is an example of these systems. In fact, while tumor cells are involved in the phase change from interphase to mitosis (the delayed event) they can also die.

\begin{figure}[t]
\begin{center}
 \includegraphics[width=11cm]{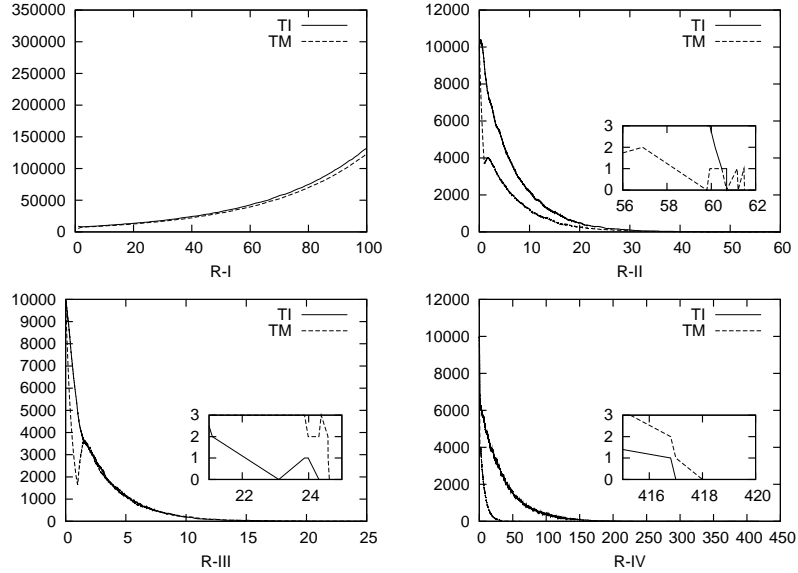}
 \caption{DDA simulation of the stochastic model with $\sigma=1$ for the regions described in
 Figure~\ref{fig:regions}. On the x-axis time is given in {\em days} and on the y-axis is given the {\em number of cells}.}\label{fig:Tau-1-DDA}
\end{center}
\end{figure}

\begin{figure}
\begin{center}
 \includegraphics[width=11cm]{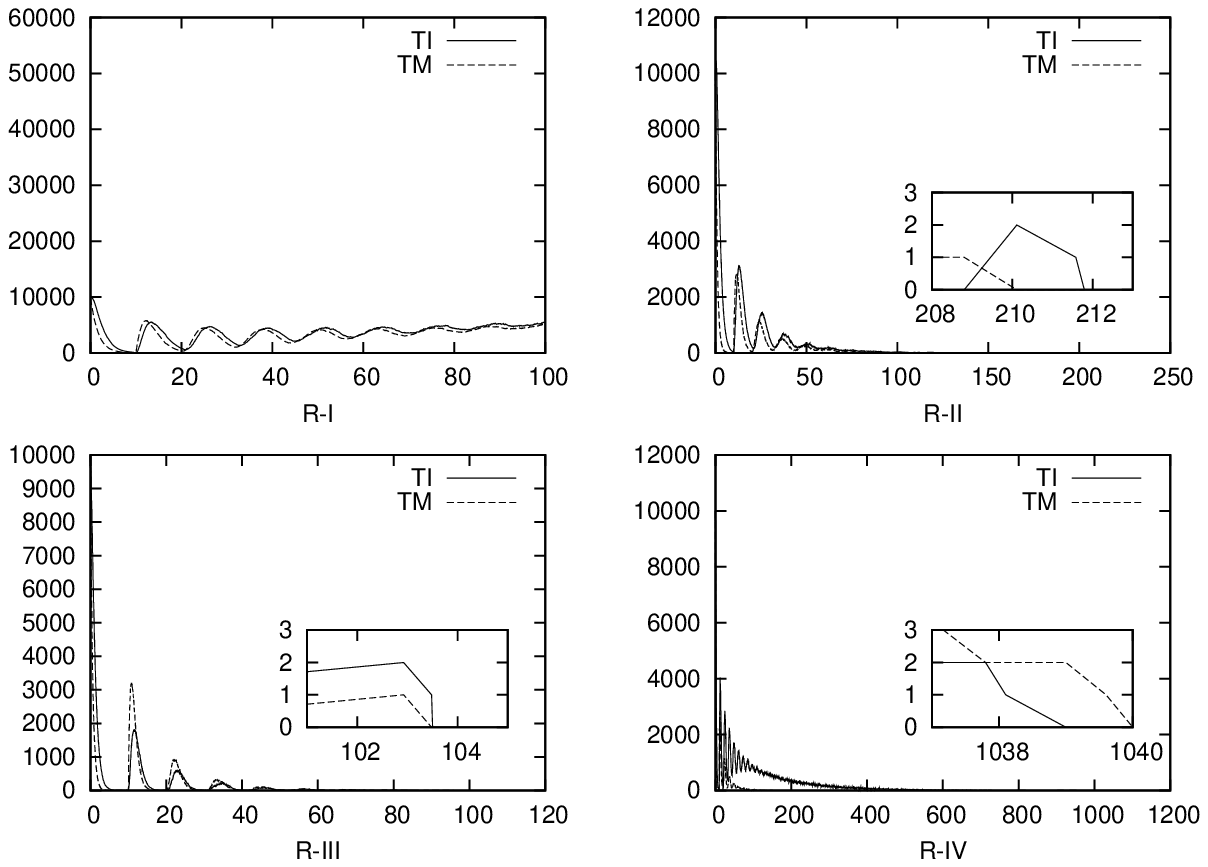}
\caption{DDA simulation of the stochastic model with $\sigma=10$ for the regions described in
 Figure~\ref{fig:regions}. On the x-axis time is given in {\em days} and on the y-axis is given the {\em number of cells}.}\label{fig:Tau-10-DDA}
\end{center}
\end{figure}

We applied the DSSA by Barrio {\em at al.} (we refer to the simulations done by applying this DSSA as DDA simulations) to a chemical reaction model corresponding to the DDE model of tumor growth recalled in Section~\ref{sect:dde-model}. The
reactions of the model are the following:
\begin{gather*}
T_I \xrightarrow{a_1} T_M \mbox{ with delay $\sigma$} \qquad T_M \xrightarrow{a_4}
2T_I \qquad
T_I \xrightarrow{d_2} \qquad T_M \xrightarrow{d_3}\, .
\end{gather*}
We have run 100 simulations for each considered parameter setting. The
results of simulations with the same parameters as those considered in
Figures~\ref{fig:DDE-1} and \ref{fig:DDE-10} are shown in
Figures~\ref{fig:Tau-1-DDA} and \ref{fig:Tau-10-DDA}, respectively. Actually,
in the figures we show the result of one randomly chosen simulation run for each
parameter setting.

Qualitatively, results obtained with DDA simulations are the same as those
obtained with numerical simulation of the DDEs: we have exponential tumor growth
in region R-I, tumor decay in the other regions and oscillations arise when
the delay is increased. However, from the quantitative point of view we have
that in the DDA simulations the growth in region R-I and the decay in the other
regions are always slower than in the corresponding numerical simulation of the
DDEs. In fact, with $\sigma = 1$ by the numerical simulation of the DDEs we have
that in region R-I after 100 days both the quantities of tumor cells in
interphase and in mitotic phase are around 300000, while in the result of DDA
simulations they are around 130000. In the same conditions, but with $\sigma =
10$, in the numerical simulation of the DDEs we have about 47000 tumor cells in
mitosis and 57000 tumor cells in interphase, while in the DDA simulations we
have about 5000 and 5500 cells, respectively. As regards the other regions,
in Table~\ref{tab:eradication} the average tumor eradication times obtained with
DDA simulations are compared with those obtained with numerical simulation of the
DDEs (in this case with ``eradication'' we mean that the number of tumor cells
of both kinds is under the value 1). Again, we have that in DDA simulations
the dynamics is slower than in the numerical simulation of the DDEs. For instance, with $\sigma=10$,
in region R-IV the time needed for eradication in the DDEs is about
$41\%$ of the time needed in the DDA (440 against 1072), in region R-II
the percentage is smaller, $26\%$ (59 against 224), and, in region R-III, it reaches
$9\%$ (12 against 126). For the same regions with $\sigma=1$ these differences
are smaller but not negligible.

\subsection{A Purely Delayed Approach (PDA)}

In this section we propose a variant of the DSSA based on a different interpretation of delays, namely a Stochastic Simulation Algorithm which follows a ``purely delayed approach'' (PDA). With this interpretation we try to overcome the fact that in the DDA the reactants cannot have other interactions. Furthermore, differently from Barrio {\em et al.}, we  use the same interpretation of delays to define the method for firing both non-consuming and consuming reactions. This interpretation of delays was firstly implicitly adopted by Bratsun {\em et al.} in~\cite{BVTH05},  to model a very simple example of protein degradation.

The approach we propose consists in firing a reaction completely when its associated scheduled events is handled, namely removing its reactants and inserting its products after the delay. The fact that we simply schedule delayed reactions without immediately removing their reactants motivates the terminology of ``purely delayed''.  Notice that non-consuming reactions are handled in the same way by DDA and PDA.

\begin{figure}
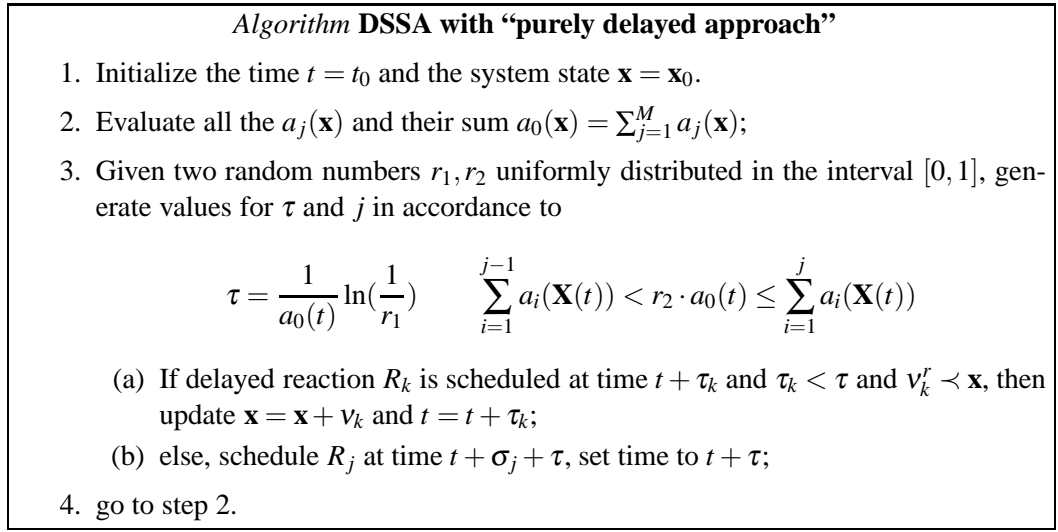

\framebox{
\begin{minipage}{0.85\textwidth}
\begin{center}
\noindent \textit{Algorithm} \textbf{DSSA with ``purely delayed approach''}
\begin{enumerate}
\item Initialize the time $t = t_0$ and the system state $\xx=\xx_0$.

\item Evaluate all the $a_j(\xx)$ and their sum $a_0(\xx)=\sum_{j=1}^Ma_j(\xx)$;

\item Given two random numbers $r_1,r_2$ uniformly distributed in the interval $[0,1]$, generate values for $\tau$ and $j$ in accordance to
{\[
 \tau = \dfrac{1}{a_0(t)} \ln(\dfrac{1}{r_1}) \qquad \sum_{i=1}^{j-1} a_i(\XX(t)) < r_2 \cdot a_0(t) \leq \sum_{i=1}^j  a_i(\XX(t))
\]}
\begin{enumerate}
 \item {If delayed reaction $R_k$ is scheduled at time $t+\tau_k$ and $\tau_k < \tau$ and $\nu_k^r \prec \xx$, then  update $\xx = \xx + \nu_k$ and $t = t + \tau_k$;}
 \item {else, schedule $R_j$ at time $t+\sigma_j+\tau$, set time to $t+\tau$};
\end{enumerate}
\item go to step $2$.
\end{enumerate}
\end{center}
\end{minipage}}
\caption{The DSSA with ``purely delayed approach''.}
\label{fig:DSSA-PDA}
\end{figure}

In this interpretation of delays it may happen that, when handling a scheduled reaction, the reactants may not be present in the current state. In fact, they could have been destroyed or transformed by other interactions happened after the scheduling. In this case, the scheduled reaction has to be ignored. To formalize this, we know that a reaction $R_j$ can be applied only if its reactants are all present in the current state of the simulation. Algebraically this corresponds to the fact that $\nu_j^r \prec \xx$ where $\nu_j^r$ is the state--change vector of the reactants of reaction $R_j$, the system is described by $\xx$ and $\prec$ is the ordering relation defined as $\forall i=1,\ldots,N.\;-\nn_{ij}^R \leq X_i(t)$. In order to verify that a scheduled reaction can effectively fire, it will be sufficient to check whether this condition holds. The formal definition of the DSSA with PDA is given in Figure~\ref{fig:DSSA-PDA}.

\begin{figure}[t]
\begin{center}
 \includegraphics[width=11cm]{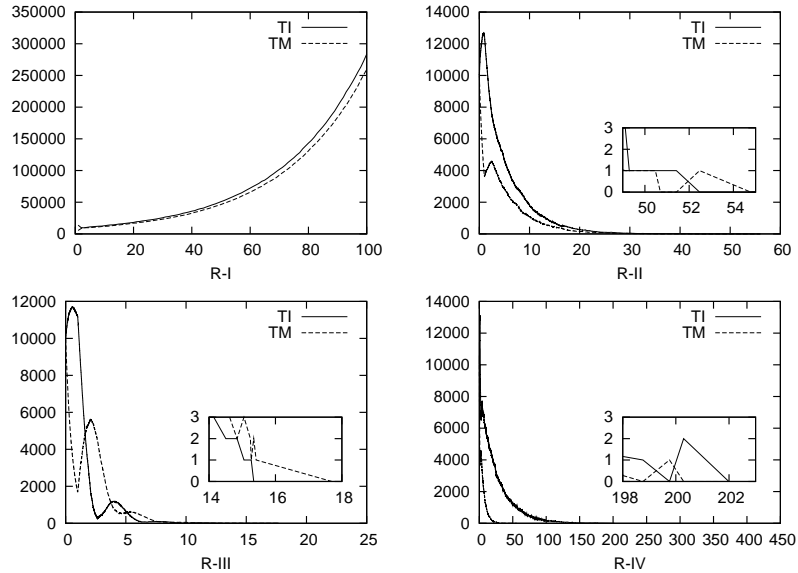}
 \caption{PDA simulation of the stochastic model with $\sigma=1$ for the regions described in
 Figure~\ref{fig:regions}. On the x-axis time is given in {\em days} and on the y-axis is given the {\em number of cells}.}\label{fig:Tau-1-PDA}
\end{center}
\end{figure}

\begin{figure}
\begin{center}
 \includegraphics[width=11cm]{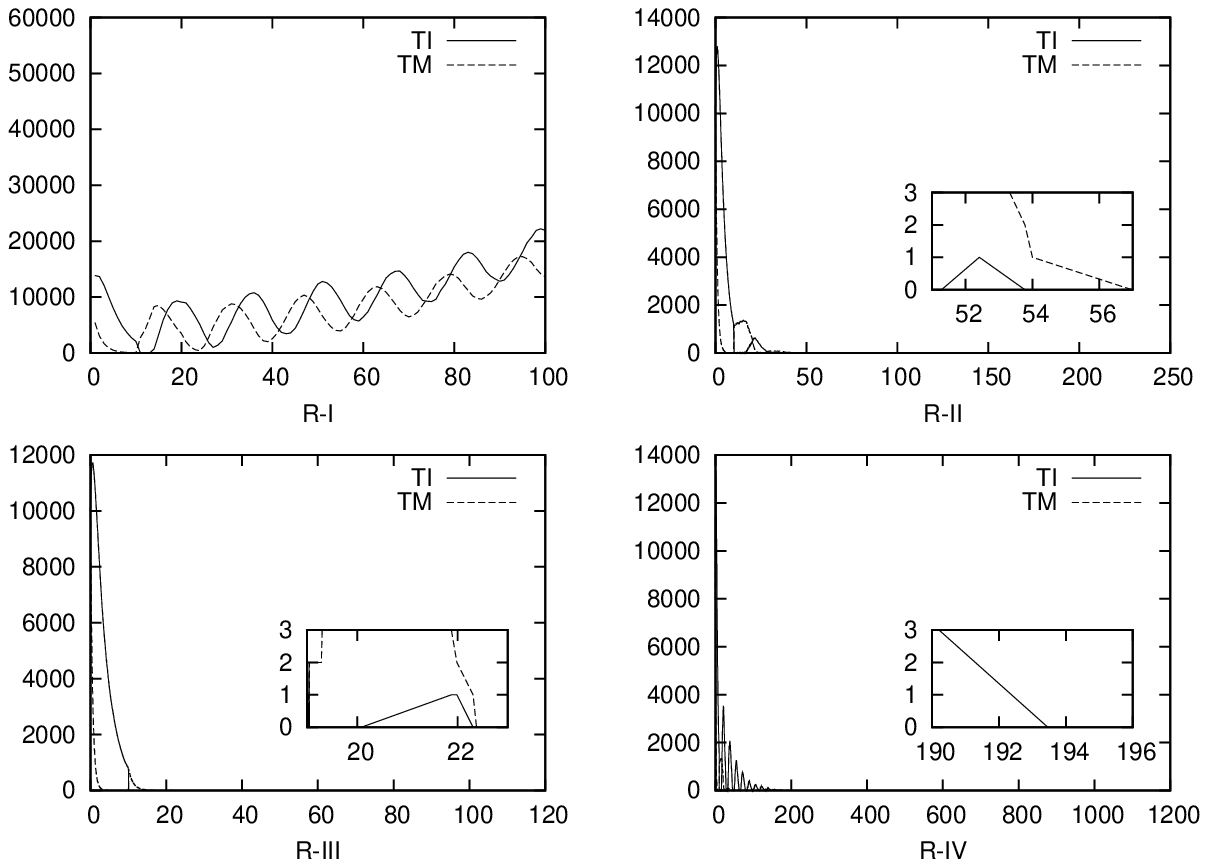}
 \caption{PDA simulation of the stochastic model with $\sigma=10$ for the regions described in
 Figure~\ref{fig:regions}. On the x-axis time is given in {\em days} and on the y-axis is given the {\em number of cells}.}\label{fig:Tau-10-PDA}
\end{center}
\end{figure}

\begin{table}[t]
\begin{center}
\begin{tabular}{l|c|c|c|}
            & \;DDEs\;     & \;DDA Simulation\;     & \;PDA Simulation\;\\
\hline
R-II with $\sigma = 1.0$    & 50    & 64            & 51        \\
R-II with $\sigma = 10.0$    & 59    & 224            & 67        \\
\hline
R-III with $\sigma = 1.0$    & 15    & 29            & 17         \\
R-III with $\sigma = 10.0$    &  12    & 126            & 20        \\
\hline
R-IV with $\sigma = 1.0$    & 238    & 302            & 214        \\
R-IV with $\sigma = 10.0$    & 440    & 1072            & 248       \\
\hline
\end{tabular}
\medskip
\caption{Average eradication times given in {\em days} for DDE model, DDA and PDA stochastic models. For the stochastic models the entries represent the sample of 100 simulations.}\label{tab:eradication}
\end{center}
\end{table}
As for the DDA, we have run 100 simulations of the stochastic model
of tumor growth for each considered parameter setting.
The results of simulations (we refer to these simulations as PDA simulations) with the same parameters as those considered in Figures~\ref{fig:DDE-1} and \ref{fig:DDE-10} are shown in
Figures~\ref{fig:Tau-1-PDA} and \ref{fig:Tau-10-PDA}, respectively. Actually,
in the figures we show the result of one randomly chosen simulation run for each
parameter setting.

Qualitatively, results obtained with PDA simulations are the same as those
obtained with numerical simulation of the DDEs (and with DDA
simulations). From the quantitative point of view we have
that in the PDA simulations the growth in region R-I with $\sigma=1$ is almost equal
to the corresponding numerical simulation of the DDEs (about 300000
tumor cells in both mitosis and interphase after 100 days, we recall that the DDA had reached values
around 130000). On the contrary, with $\sigma=10$, the difference between
DDEs and PDA is higher: we have about 22000 tumor cells in interphase against
57000 for the DDEs and 5500 for the DDA, and 16000 tumor cells in mitosis
against 47000 for the DDEs and 5000 for the DDA.

As regards the other regions, in Table~\ref{tab:eradication} the average tumor eradication times obtained with
PDA simulations are compared with those obtained with numerical simulation of the
DDEs (again, in this case with ``eradication'' we mean that the number of tumor cells
of both kinds is under the value 1). In PDA simulations the dynamics is generally slower than in the numerical simulation of the DDEs but it is faster than the DDA one. With $\sigma=10$,
in region R-IV the time needed for eradication in the PDA is
smaller than the one in the DDEs (248 days against 440, DDA is 1072). In region R-II
the values are: 67 days for the PDA and 59 days for the DDEs, DDA is 224.
In region R-III values are: 20 days for the PDA, 12 days for the DDEs, and 126 days for DDA.


It is important to remark that differences between delay stochastic simulation results and numerical solutions of DDEs are also influenced by the initial conditions. The numerical solution of the DDEs assumes the initial population to be constant and greater than zero in the time interval $[-\sigma,0]$. This allows delayed event to be enabled in the time interval $[0,\sigma]$. Both variants of the DSSA start to schedule delayed events from time 0, hence delayed reactions can fire only after the time $\sigma$. This result, when $\sigma$ is great enough, in a behaviour that is, in general, delayed with respect to that given by the DDEs.

\section{Discussion}

In the previous sections we showed two different approaches to the firing of delayed reactions. The two approaches can be conveniently used for dealing with two different classes of delayed reactions.
The delay as duration approach suitably deals with reactions in which reactants cannot participate, whenever scheduled, in other reactions. On the other hand, the purely delayed approach can be conveniently used in cases in which reactants can be involved in other reactions during the delay time.

These two different notions of delay have been presented in the framework of Petri nets with time information. In particular, in Timed nets~\cite{Ramchandani} a notion of delay similar to a duration appears; differently, in Time nets~\cite{Heiner} the notion of delay corresponds to our purely delayed approach.

In the example we have shown, cells in the interphase, which wait for entering the mitotic phase, can be involved in another reaction, namely their death. Thus in this example the purely delayed approach seems to be more appropriate for capturing the behaviour of the real system. Obviously, there are biological systems in which, due to the heterogeneity of reactions, both the approaches should be used and we plan to investigate, in the future, the possibility of combining the two approaches in a unique framework.

\bibliographystyle{eptcs} 

\end{document}